\documentclass[sigconf]{acmart}
\AtBeginDocument{%
  }

\setcopyright{acmlicensed}
\copyrightyear{2025}
\acmYear{2025}
\setcctype{by}
\acmConference[CIKM '25]{Proceedings of the 34th ACM International Conference on Information and Knowledge Management}{November 10--14, 2025}{Seoul, Republic of Korea}
\acmBooktitle{Proceedings of the 34th ACM International Conference on Information and Knowledge Management (CIKM '25), November 10--14, 2025, Seoul, Republic of Korea}
\acmISBN{978-1-4503-XXXX-X/2018/06}
\acmDOI{XXXXXXX.XXXXXXX}

\usepackage{color}
\usepackage{url}
\usepackage{algorithm}
\usepackage{algpseudocode} 
\usepackage{multirow}

\usepackage{hhline} 
\usepackage{subcaption}

\begin{document}

\title{Contrastive Multi-View Graph Hashing}



\author{Yang Xu}
\affiliation{%
  \department{State Key Laboratory for Novel Software Technology \& School of Artificial Intelligence}
  \institution{Nanjing University}
  \city{Nanjing}
  \country{China}
}
\email{xuyang@lamda.nju.edu.cn}

\author{Zuliang Yang}
\affiliation{%
  \department{State Key Laboratory for Novel Software Technology \& School of Artificial Intelligence}
  \institution{Nanjing University}
  \city{Nanjing}
  \country{China}
}
\email{yangzl@lamda.nju.edu.cn}

\author{Kai Ming Ting}
\affiliation{%
  \department{State Key Laboratory for Novel Software Technology \& School of Artificial Intelligence}
  \institution{Nanjing University}
  \city{Nanjing}
  \country{China}
}
\email{tingkm@nju.edu.cn}




\begin{abstract}
  Multi-view graph data, which both captures node attributes and rich relational information from diverse sources, is becoming increasingly prevalent in various domains. The effective and efficient retrieval of such data is an important task. 
  Although multi-view hashing techniques have offered a paradigm for fusing diverse information into compact binary codes, they typically assume attributes-based inputs per view. This makes them unsuitable for multi-view graph data, where effectively encoding and fusing complex topological information from multiple heterogeneous graph views to generate unified binary embeddings remains a significant challenge. In this work, we propose Contrastive Multi-view Graph Hashing (CMGHash), a novel end-to-end framework designed to learn unified and discriminative binary embeddings from multi-view graph data. CMGHash learns a consensus node representation space using a contrastive multi-view graph loss, which aims to pull $k$-nearest neighbors from all graphs closer while pushing away negative pairs, i.e., non-neighbor nodes. Moreover, we impose binarization constraints on this consensus space, enabling its conversion to a corresponding binary embedding space at minimal cost. Extensive experiments on several benchmark datasets demonstrate that CMGHash significantly outperforms existing approaches in terms of retrieval accuracy.
\end{abstract}

\ccsdesc[500]{Information systems~Retrieval tasks and goals~Similarity search}
\ccsdesc[500]{Computing methodologies~Machine learning}
\ccsdesc[300]{Information systems~Graph-based database models}


\keywords{Multi-View Graph Data, Semantic Hashing, Contrastive Learning}


\maketitle

\section{Introduction}
\label{sec:introduction}

An attributed graph, which contains both node attributes and edges representing the pairwise relations between nodes, is a fundamental structure for modeling relational data~\cite{zhou2009graphintro1}. It finds wide applicability in real-world scenarios, including social networks, sensor networks, and biological networks~\cite{dong2023graphintro2,liao2018attributedintro2,pavlopoulos2011usingintro2}. For example, in a social network, users possess their own characteristics (node attributes), and their social relationships form a graph structure. Most studies in graph learning focus on first constructing effective graph representations from such an attributed graph, typically assuming a single set of node attributes and a single graph structure. These representations are then applied to various downstream tasks, such as link prediction, classification, and clustering~\cite{cai2021lineintro3,lee2018graphintro3,akbas2017attributedintro3,kipf2017semi}. While these approaches can effectively operate on what is essentially single-view graph data, they are not suitable to deal with the complexities of multi-view graph data.

\begin{figure}[t]
    \centering
    \includegraphics[width=1\linewidth]{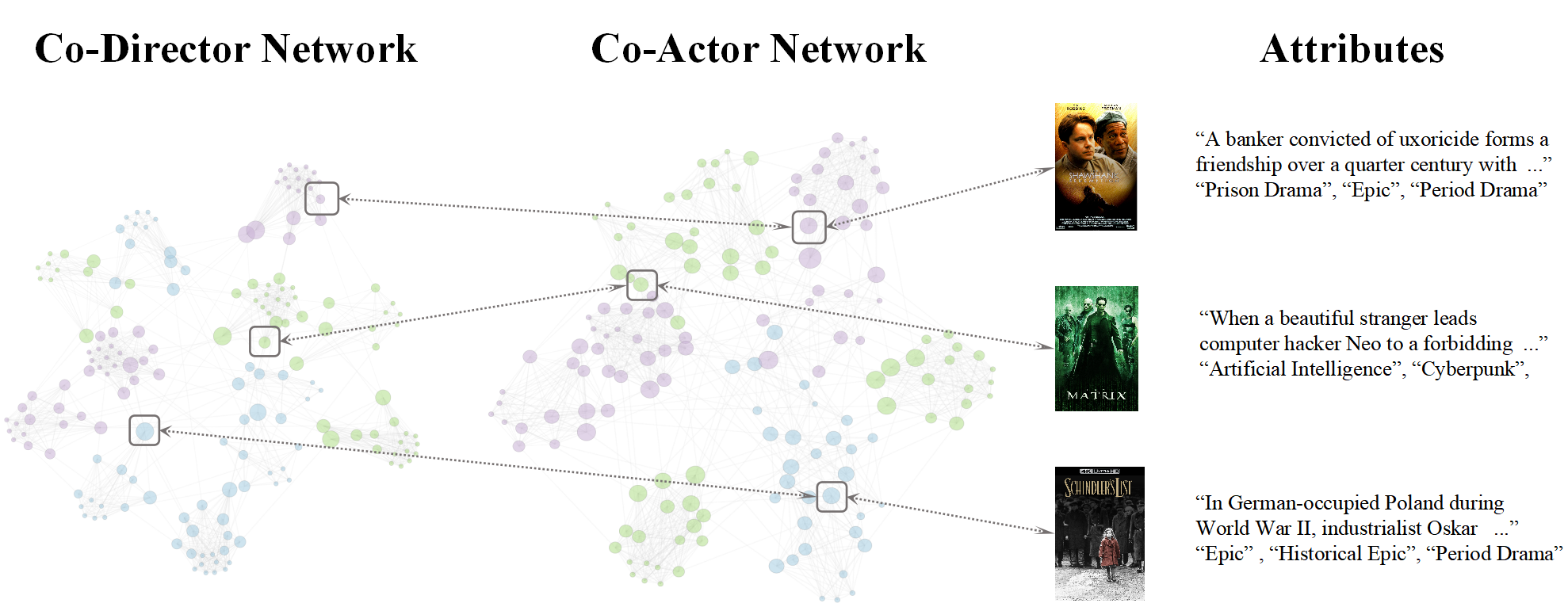}
    \caption{An illustrative example of multi-view graph data. Here, entities (e.g., movies, represented by nodes) are characterized through distinct views: a "Co-Director Network" capturing relationships based on shared directors, a "Co-Actor Network" detailing connections through common actors, and an "Attributes" view providing descriptive features. Each view offers a unique structural or feature-based perspective on the same underlying set of entities.}
    \Description{This image illustrates multi-view graph data using movies as entities (nodes). Three distinct views are presented: "Co-Director Network" shows connections between movies sharing directors, "Co-Actor Network" displays links between movies with common actors, and "Attributes" provides feature-based information for each movie. Each view offers a unique perspective on the same set of movie entities.
    \textcolor{red}{The font size is too tiny in the `Attributes' column}
    }
    \label{fig:movie}
\end{figure}

Real-world graph data, obtained from various extractors or collected from multiple sources, are naturally characterized by different features or views~\cite{lin2021graph,li2025clustering,pan2021multi}. In other words, nodes could have various attributes or interact with each other through multiple types of relationships (see the movie network example in Figure~\ref{fig:movie}). Although fusing the rich information embedded in multi-view graph data has yielded impressive successes across a variety of data mining tasks~\cite{ma2018drug,khan2019multi,geng2019spatiotemporal}, effectively and efficiently retrieving such data remains a significant challenge. This is largely due to the intrinsic complexity of graph properties, especially considering the critical need to integrate information from multiple graph views~\cite{cai2018comprehensive}. 
Existing representation learning methods often learn high-dimensional consensus representations. In some cases, particularly when explicitly modeling pairwise similarities or using certain spectral techniques, this can involve structures as large as $N\times N$ (where $N$ is the number of nodes)~\cite{pan2021multi,nie2017self,yu2023multi,ling2023dual}. As the number of nodes readily scales to hundreds of thousands or beyond, the prohibitive storage cost and computational overhead of these high-dimensional continuous representations become a critical bottleneck for efficient retrieval~\cite{luo2023survey}.

Multi-view hashing (MVH) has emerged as a promising paradigm to address the challenges of high dimensionality and computational overhead. MVH techniques aim to learn compact binary codes that fuse information from multiple data sources, thereby enabling significant data compression and accelerated retrieval speeds~\cite{zhu2023multi,hu2024cross}. Specifically, similarity search is reduced to extremely fast bitwise XOR operations in the Hamming space, making it highly suitable for large-scale database and real-time applications~\cite{wang2017survey,luo2023survey}.
MVH has shown considerable success in various domains from image and multimedia retrieval to more complex cross-modal applications, where data are often represented by multiple sets of features~\cite{liang2024multi,li2025deep,liu2025distribution}. However, a common assumption in these studies is that each view consists of 
vector-based attribute features, which are not suited for representing topological graph data. As a result, directly applying these existing MVH methods to multi-view graph data, where individual views can themselves be complex graph structures, could yield suboptimal performance due to their inability to adequately capture and leverage the intricate structural properties inherent in and across the graph views. This underscores the pressing need to develop specialized hashing methods capable of effectively exploiting the latent characteristics of data represented by multiple graph topologies.

To address this, we introduce Contrastive Multi-view Graph Hashing (CMGHash), a novel end-to-end framework 
for learning discriminative binary embeddings from multi-view graph data. CMGHash strategically leverages the rich information embedded within graph structures across multiple views. Conceptually, it first uses a graph filtering technique to obtain smoothed node representations for each view. These representations are then embedded into a unified low-dimensional consensus space, the learning of which is guided by a contrastive loss specifically designed for multi-view graphs. This loss aims to maximize the agreement between positive pairs, that is, an anchor node and its neighbors across different views, while distinguishing them from negative pairs. 
Moreover, binarization constraints are imposed on this learned consensus space to facilitate its conversion into compact binary codes with minimal information loss.

The main contributions of our work are summarized as follows:
\begin{itemize}
    \item We propose CMGHash, a novel end-to-end contrastive hashing framework specifically designed for multi-view graph data. To the best of our knowledge, this the first attempt to address the generation of binary codes from data characterized by multiple graph topologies.
    \item We evaluated the performance of existing approaches in learning binary embeddings from multi-view graph data, including single-view graph hashing, multi-view graph learning, and multi-view hashing, uncovering their limitations.
    \item Extensive experiments conducted on five benchmark multi-view graph datasets show that our proposed CMGHash significantly outperforms existing approaches in terms of retrieval accuracy.
\end{itemize}

\section{Related Work}
\label{sec:related_work}

\subsection{Graph Representation Learning}

Graph representation learning is a cornerstone for analyzing graph-structured data, aiming to embed graph components, typically nodes, into a consensus representation space while preserving both attribute and structural information. Existing approaches can be broadly categorized into matrix factorization-based techniques, which learn embeddings by decomposing matrix representations of graphs (\textit{e.g.}, adjacency or Laplacian matrices)~\cite{brochier2019globalx21,ahmed2013distributedx21}; random walk-based methods, which leverage techniques like Skip-gram on node sequences generated from random walks to capture neighborhood information~\cite{perozzi2014deepwalkx21,grover2016node2vecx21}; and more recently, graph neural networks (GNNs), which employ GNNs to generate embeddings~\cite{velivckovic2018graph,beaini2021directionalx21,kipf2017semi}. Nevertheless, these techniques only focus on single-view graph data (a single set of node attributes and a single graph structure), and primarily produce continuous, real-valued embeddings.

\textbf{Multi-view graph representation learning.} Real-world data often come from various sources, forming multi-view graph data that include multiple attribute sets or varied graph relations. The goal of multi-view graph representation learning is to integrate such data to derive more powerful consensus representations~\cite{lin2021graph,li2025clustering,xiao2024graph,du2025mvf}. Existing methods can be broadly divided into three categories: multi-attribute learning~\cite{zheng2021multiMAL,hou2022shiftingMAL,mai2020modalityMAL}, multi-relation learning~\cite{xu2023mvhgnMRL,lu2021predictingMRL,wang2023learningMRL}, and mixed learning~\cite{LHGN,xu2023simpleMML,wang2023consistentMML,pan2021multi,yu2023multi}. For example, LHGN~\cite{LHGN} is a latent heterogeneous graph network, which first encodes information from different views into a unified latent representation. Then, it constructs a heterogeneous graph based on this unified latent representation and applies graph learning to integrate the structural information between samples. An aggregation network then fuses the features learned from different views.

Although these multi-view graph representation learning methods achieve impressive performance in learning continuous embeddings, they often produce high-dimensional representations or correlation matrix not directly suitable for large-scale, rapid similarity search due to computational and storage costs. This motivates the exploration of hashing techniques specifically for such multi-view graph representations, which remains an under-explored area. For example, MCGC is specifically tailored for multi-view graph clustering~\cite{pan2021multi}. Its goal is to obtain a unified correlation matrix that is optimal for a subsequent spectral clustering task, rather than producing compact node representations for efficient retrieval.

\textbf{Graph hashing.} 
In parallel, graph hashing has been explored to generate compact binary codes for nodes, facilitating efficient similarity search and reduced storage overhead in large-scale graph applications~\cite{zhou2018graph,liu2011hashingGHL,qin2020ghashingGHL,yeh2022embeddingGHL,jiang2015scalableGHL,tan2020learning}. A naive approach is to first transform the graph into continuous representations, and then apply established hashing techniques~\cite{luo2023survey,wang2017survey} on these continuous representations to obtain the final binary codes. However, this faces the dual information loss problem. First, employing graph representation learning to embed continuous representations could lead to some information loss. Second, the hashing step itself introduces further information loss by binarizing these representations. Therefore, graph hashing methods must overcome the challenge of preserving the most important structural and attribute similarities in the final binary codes while minimizing losses from both representation and binarization. 
For example, HashGNN~\cite{tan2020learning}, a framework designed to learn discrete hash codes directly from graph data, such as user-item interaction networks, primarily for efficient recommender systems, rather than focusing on the fusion of multi-view graph data. The method utilizes a GNN encoder to learn continuous node representations from the input graph. These learned representations are subsequently transformed into binary hash codes by a dedicated hash layer, enabling fast similarity search in the Hamming space
However, existing methods are only applicable to single-view graph data input.

\subsection{Multi-view hashing}

Multi-view hashing is an emerging popular paradigm, which has received widespread research attention in multimedia and cross-modal retrieval by effectively embedding data from different views or modalities into binary embeddings~\cite{liang2024multi,liu2025distribution,shen2018multiview,yan2020deep,hu2025unsupervised}. Similar to graph hashing, multi-view hashing also faces the challenge of reducing the dual loss of fusing multi-view representations and binarization~\cite{zhu2023multi,hu2024cross}. For example, D-MVE-Hash~\cite{yan2020deep}, a supervised framework for image retrieval that enhances hashing performance by fusing multiple feature views of images. It achieves this fusion by first computing a view-relation matrix derived from evaluating the stability and relationships among these diverse image features; this matrix subsequently guides the overall network optimization. The framework then employs specific fusion methods, such as replication fusion, view-code fusion, or probability view pooling, to integrate basic binary codes with multi-view binary codes in the Hamming space, guided by the learned view-relation matrix.

It is worth noting that existing methods assume that data from different views represent a portion of the attributes from different sources, and their goal is to fuse these attributes to obtain consensus binary embeddings. Therefore, these methods are not suitable for multi-view graph data, which simultaneously contains multiple sets of attributes as well as multiple relational graphs. Since these relational graphs contain rich topological structure information, simply treating them as attributes could lead to suboptimal performance.

Despite the advances in both graph hashing and multi-view hashing, research specifically addressing the unique challenges of hashing multi-view graph data, where each view can itself be a graph with distinct topology and attributes, remains nascent. To the best of our knowledge, a dedicated end-to-end framework for this specific task is still lacking. This motivates us to develop Contrastive Multi-view Graph Hashing (CMGHash), the first specialized hashing framework for multi-view graph data, which effectively integrates multiple attribute sets and diverse graph topologies to achieve better binary embeddings.

\section{Problem Definition}
\label{sec:problemdef}

 A summary of the key notations and parameters used in this paper is concluded in Table \ref{tab:notations}.

\begin{table}[t!]
\centering
\caption{Key notations and descriptions used in this paper.}
\label{tab:notations}
\resizebox{\columnwidth}{!}{
\begin{tabular}{l|l}
\toprule
\textbf{Notation} & \textbf{Description} \\
\midrule
$N$ & Number of nodes. \\
$V$ & Number of views. \\
$X^{(v)} \in \mathbb{R}^{N \times d^{(v)}}$ & Attribute matrix of view $v$. \\
$A^{(v)} \in \mathbb{R}^{N \times N}$ & Adjacency matrix of view $v$. \\
$L^{(v)} \in \mathbb{R}^{N \times N}$ & Normalized graph Laplacian matrix of view $v$. \\
$S^{(v)} \in \mathbb{R}^{N \times d^{(v)}}$ & Smoothed feature matrix of view $v$. \\
$K$ & Length of the binary hash codes. \\
$U \in \mathbb{R}^{N \times K}$ & Continuous consensus representations. \\
$\mathbf{B} \in \{-1, 1\}^{N \times K}$ & Consensus binary embeddings. \\
$m, s$ & Parameters for graph filtering. \\
$\mathcal{L}_{C}$ & Contrastive multi-view graph loss. \\
$\mathcal{L}_{Q}$ & Quantization loss. \\
$\mathcal{L}_{BB}$ & Bit balance loss. \\
$\tau$ & Temperature parameter in the contrastive loss. \\
$\alpha, \beta, \eta$ & Balance parameters for different loss terms. \\
\bottomrule
\end{tabular}}
\end{table}

Formally, a single-view graph is often represented as $G=(\mathcal{V}, E)$, where $\mathcal{V}$ denotes the set of nodes, and $E \subseteq \mathcal{V} \times \mathcal{V}$ represents the set of edges indicating relationships between these nodes.
When node attributes are also considered, it is often referred to as an attributed graph, $G=(\mathcal{V}, X, E)$, where $X$ is a matrix of node features.
This foundational definition serves many graph-based learning tasks. 
As real-world entities and their interconnections are frequently multifaceted, the multi-view graph data can provide a richer representation by incorporating information from multiple sources or perspectives, representing different types of relationships or various sets of node characteristics. Formally, the multi-view graph data $\mathcal{G}$ are defined as follows:

\begin{definition} (\textit{Multi-view graph data})
 A given multi-view graph data $\mathcal{G}$ is defined as $\mathcal{G} = \{\mathcal{V},\mathcal{X},\mathcal{E}\}$, where $\mathcal{V}$ denotes the sets of $N$ nodes, $\mathcal{X}=\{X^{(v)}\}_{v=1}^{V}$ represents the attribute matrices  in $V$ views, and $\mathcal{E}=\{E^{(v)}\}_{v=1}^{V}$ represents the node relations of edges in $V$ views, with $V$ being the total number of views. ${X}^{(v)} = \{x_1^{(v)},\dots,x_N^{(v)}\}^{\top}$ is the attribute matrix of view $v$ {having $N$ nodes}, and $e_{ij}\in {E}^{(v)}$ is the relationship between nodes $i$ and $j$ in view $v$.
\end{definition}

Here we focus on learning a model/function $f_{\Theta}$ which transforms multi-view graph data into binary embeddings. The problem of multi-view graph hashing can thus be formulated as:

\begin{definition} (\textit{Multi-view graph hashing}) Given multi-view graph data $\mathcal{G} = \{\mathcal{V},\mathcal{X},\mathcal{E}\}$, the  objective of multi-view graph hashing is to simultaneously explore both sets of attributes and node relations: $f_{\Theta}(\mathcal{X},\mathcal{E})\rightarrow \mathbf{B}$, where $\Theta$ is the trainable parameters of the model, and $\mathbf{B}\in\{-1,1\}^{N\times K}$ is the matrix of learned binary embeddings, where each row $B_i$ represents the $K$-bit code for node $i$.
\end{definition}

These binary codes $\mathbf{B}$ are sought to satisfy the similarity preservation property that the Hamming distance $d_H(B_i, B_j)$ between the binary codes of any two nodes $i$ and $j$ should effectively approximate their underlying semantic similarity as captured by the original multi-view graph $\mathcal{G}$.
Specifically, nodes considered similar across the multiple views, integrating both structural and attribute information, should be mapped to binary codes that are close in Hamming distance, while dissimilar nodes should be mapped to binary codes that are far apart.

\section{Methodology}
\label{sec:methodology}

\begin{figure*}[t]
    \centering
    \includegraphics[width=0.92\linewidth]{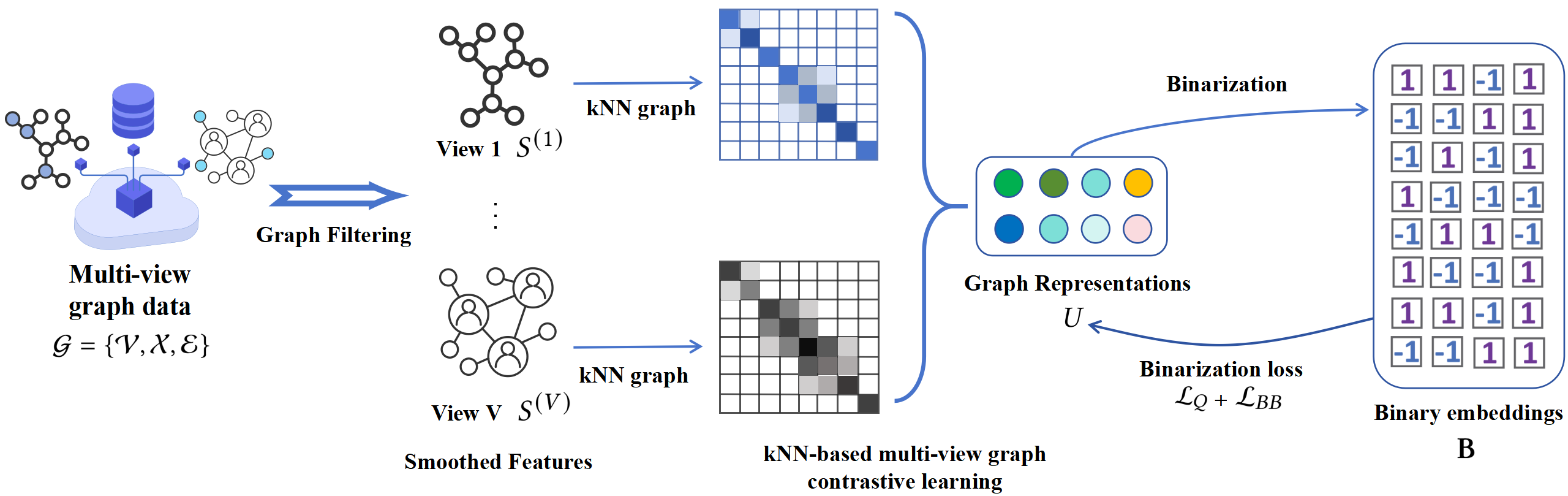}
    \caption{Architecture of the proposed CMGHash method. The framework begins with multi-view graph data $\mathcal{G}=\{\mathcal{V},\mathcal{X},\mathcal{E}\}$, where each view's features undergo graph filtering to produce smoothed features ($S^{(v)}$). These smoothed features are then processed to learn unified continuous graph representations ($U$). The learning of $U$ is guided by two main components: a KNN-based multi-view graph contrastive learning mechanism, and binarization losses ($\mathcal{L}_Q + \mathcal{L}_{BB}$). Finally, the learned continuous representation $U$ is transformed into binary codes ($\mathbf{B}$).}
    \Description{A diagram illustrating the architecture of the CMGHash method. The process begins with multi-view graph data as input, which then undergoes graph filtering to produce smoothed features. These features are subsequently used to learn a unified continuous graph representation, denoted as $U$. This learning stage is guided by two parallel components: a KNN-based multi-view graph contrastive learning mechanism and a binarization loss component. In the final stage, the continuous representation $U$ is transformed to produce the final output: the binary hash codes, denoted as $\mathbf{B}$.}
    \label{fig:method}
\end{figure*}

Inspired by the successful application of contrastive learning in multi-view graph clustering tasks~\cite{pan2021multi}, we have introduced this approach to multi-view graph hashing. Specifically, we propose Contrastive Multi-view Graph Hashing (CMGHash), a novel
end-to-end framework designed to learn unified and discriminative binary codes from multi-view graph data. The entire process of the method is illustrated in Figure~\ref{fig:method}.
CMGHash first employs graph filtering on node attributes from each view to denoise the data and capture essential structural information in smoothed representations. Subsequently, it uses a $k$NN-based contrastive loss function to guide the learning of the consensus representations. Finally, the binarization constraints are added to the learning objective to facilitate their conversion into binary embeddings with minimal transformation loss.

\subsection{Representation Smoothing via Graph Filtering}
\label{subsec:graph_filtering}

Originating from multiple sources, each view of a given multi-view graph data could be noisy and incomplete, but important factors, such as geometry and semantics, tend to be shared among all views. To filter out noise, an effective approach is graph filtering, which reduces noise and achieves smoothed representations, thereby improving the performance of downstream tasks~\cite{peng2024powerful}. Given an attribute matrix $X^{(v)} \in \mathbb{R}^{N \times d^{(v)}}$, its smoothed representations $S^{(v)}$ can be obtained by solving the following optimization problem~\cite{zhu2021interpreting,pan2021multi}
\begin{equation}
    \min_{S^{(v)}} \|S^{(v)} - X^{(v)}\|_{F}^{2} + s \operatorname{Tr}({S^{(v)}}^{\top}{L^{(v)}}{S^{(v)}}),
\label{eq:GF1}
\end{equation}
where $s$ is a positive balance parameter and $L^{(v)}$ is the Laplacian matrix associated with $X^{(v)}$. Taking the derivative of Eq.~(\ref{eq:GF1}) w.r.t. ${S^{(v)}}$ and setting it to zero, we can obtain ${S^{(v)}}$ by
\begin{equation}
    {S^{(v)}} = (I + s{L^{(v)}})^{-1}{X^{(v)}}.
\end{equation}

For simplicity, we can approximate $H^{(v)}$ by its $m$-th order Taylor series expansion, \textit{i.e.},
\begin{equation}
    {S^{(v)}} = (I - s{L^{(v)}})^{m}{X^{(v)}},
\label{eq:smooth}
\end{equation}
where $m$ is a non-negative integer. Through graph filtering, undesirable high-frequency noise can be filtered out while all the geometric features of the graph are preserved. This process yields $\{S^{(1)},\dots,S^{(V)}\}$, which serves as the input for the following multi-view graph hashing.

\subsection{Multi-view graph hashing}
\label{subsec:learning_codes}

Directly fusing multi-view graph data into low-dimensional consensus binary embeddings faces significant challenges because each bit can only take a value of $-1$ or $1$. Consequently, even subtle adjustments can greatly impact the 
model learning, making it difficult to directly learn the optimal binary representation. 
Our method is to learn a continuous representation, $U \in \mathbb{R}^{N \times K}$, whose dimensionality is consistent with the bit size of the final learned binary representation $\mathbf{B} \in \{-1, 1\}^{N \times K}$. 
For the learned $U$, we constrain each of its values to to close to either $-1$ or $1$, thereby ensuring that the capability of the binary representation $\mathbf{B}$ obtained from $U$ is almost identical to that of $U$.

The objective function $\mathcal{J}$ of CMGHash is formulated as
\begin{gather}
     \mathcal{J} = \min_{U, \lambda^{(v)}} \underbrace{
     \sum_{v=1}^{V}\lambda^{(v)} \mathcal{L}_{C}^{(v)}(U) + \eta \sum_{v=1}^{V} (\lambda^{(v)})^{\gamma}}_{\text{Contrastive}\; \text{terms}} + \underbrace{ \alpha \mathcal{L}_{Q}(U) + \beta \mathcal{L}_{BB}(U)}_{\text{Binarization}\; \text{terms}},
\label{eq:J}
\end{gather}
where $\alpha, \beta$ and $\eta$ are hyperparameters that balance the contribution of each loss component, $\gamma$ is a smooth parameter, and $\lambda^{(v)}$ is the weight of view $v$. This objective function consists of the following key components: contrastive multi-view graph loss $\mathcal{L}_C$, adaptive view weighting regularizer $\sum_{v=1}^{V} (\lambda^{(v)})^{\gamma}$, quantization loss $\mathcal{L}_Q$, and bit balance loss $\mathcal{L}_{BB}$.

\subsubsection{\textbf{Contrastive Multi-view Graph Loss} $\mathcal{L}_C$}
\label{subsec:contrastive_loss}

Contrastive learning is typically operated at the instance-level. 
For multi-view data, considering the representations of the same instance in different views as positive pairs is a foundational strategy that reinforces their shared identity. However, this approach focuses only on instance-level consistency and does not fully exploit the rich, view-specific structural semantics embedded in each graph's topology. For example, in a movie network, a node's "Co-Author" neighbors define a different type of relationship than its "Co-Subject" neighbors. To capture and fuse these diverse structural patterns, we use a $k$NN-based contrastive strategy, which uniformly pull the $k$-nearest neighbors of any given node from all its views closer in the consensus space. This encourages the model to learn a more comprehensive representation that preserves the unique neighborhood context from each view.
Here, instead of learning a consensus graph representation, we use this idea to guide the learned low-dimensional representation $U$. 
Our contrastive multi-view graph loss $\mathcal{L}_C^{(v)}$ in view $v$ is formulated as
\begin{equation}
    \mathcal{L}_{C}^{(v)}(U) = -\sum_{i=1}^{N}\sum_{j \in \mathbb{N}_i^v} \log \frac{\exp(\text{sim}(U_i, U_j)/\tau)}{\sum_{p \neq i}^{N} \exp(\text{sim}(U_i, U_p)/\tau)},
\label{eq:Lc}
\end{equation}
where $U_i$ is the $K$-dimensional continuous embedding of node $i$, $\mathbb{N}_i^v$ denotes the $k$NN of node $i$ in view $v$, $\text{sim}(\cdot, \cdot)$ is a similarity function (\textit{e.g.}, cosine similarity or dot product) between two embedding vectors, and $\tau$ is a temperature parameter, which determines the sharpness of the probability distribution of positive pairs.

Since different views may have different importance or reliability to the learned representation, we adopt adaptive weighting coefficients to balance the losses from different views, which is 
\begin{equation}
    \sum_{v=1}^{V} \lambda^{(v)} \mathcal{L}_{C}^{(v)}(U).
\end{equation}

Thus, the contrastive multi-view graph loss $\sum_{v=1}^{V} \lambda^{(v)} \mathcal{L}_{C}^{(v)}(U)$ shapes the representation space $U$ by encouraging an anchor node and its $k$NN from all views to be close, while pushing it away from other nodes. This process is designed to learn a unified and discriminative low-dimensional representation that captures the structural relationships present across the different graph views.

\subsubsection{\textbf{Adaptive view weighting regularizer} $\sum_{v=1}^{V} (\lambda^{(v)})^{\gamma}$} 
\label{subsec:adaptive_weighting}

Using the adaptive view weighting mechanism, views that are more consistent with emerging shared patterns or provide clearer structural information are expected to receive higher weights. However, unconstrained learning might lead to trivial solutions, \textit{i.e.}, completely favoring one view while ignoring the contributions of others.
To ensure these weights are well-behaved (e.g., to prevent some weights from becoming excessively large or others diminishing to zero), a regularization term, specifically $\sum_{v=1}^{V}(\lambda^{(v)})^{\gamma}$, is commonly combined, where $\gamma$ is a smooth parameter.
This regularizer promotes a more distributed and stable assignment of importance across the views, discouraging degenerate solutions.

\subsubsection{\textbf{Quantization Loss} $\mathcal{L}_Q$} Recall that our goal is to obtain binary embeddings that effectively approximate the underlying semantic similarity in multi-view graph data. However, through two contrastive terms, we only obtain a continuous representation $U$. Here, we explicitly use a quantization loss, which measures the error incurred when transforming $U$ into its corresponding binary codes. Minimizing this error encourages the learned values in $U$ to approach either $-1$ or $1$, so that the final binary embeddings can better preserve the consensus information captured by $U$. The quantization loss $\mathcal{L}_Q$ is formulated as 
\begin{equation}
    \mathcal{L}_Q(U) = \sum_{i=1}^{N} \sum_{k=1}^{K} (U_{ik} - \text{sign}(U_{ik}))^2,
\label{eq:Lq}
\end{equation}
where sign($\cdot$) is the binarization operator, \textit{i.e.}, sign$(x)=1$ if $x\geq 0$ and $-1$ if $x<0$. The quantization loss then penalizes the squared difference between the continuous value and this target binary state, effectively pushing the continuous representation towards a state that is ready for binarization with minimal information loss.

\subsubsection{\textbf{Bit Balance Loss} $\mathcal{L}_{BB}$}

In existing hashing literature, achieving bit balance is widely recognized as a potential requirement for realizing superior binary embeddings, as bit-balanced representations can encode richer information~\cite{wang2017survey,luo2023survey}. For multi-view graph hashing, we find that bit balance is also an effective constraint. The bit balance loss $\mathcal{L}_{BB}$ is formulated as
\begin{equation}
    \mathcal{L}_{BB}(U) = \sum_{k=1}^{K} \left( \sum_{i=1}^{N} U_{ik} \right)^2.
\label{eq:Lbb}
\end{equation}
This loss encourages the bits in the learned hash codes to be uniformly distributed.

In summary, the objective function $\mathcal{J}$ of CMGHash, by integrating these four loss components, transforms a given multi-view graph data of $N$ nodes into an $N\times K$ matrix of $N$ binary embeddings that effectively preserve the semantic information in the original data. Specifically, the contrastive terms utilize the $k$NN neighborhood structure from multiple graphs to learn a low-dimensional continuous consensus representation $U$, and through binarization terms, this low-dimensional consensus representation can be transformed into corresponding binary embeddings with minimal information cost.

\subsection{Optimization}
\label{sec:optimization}

The objective function $\mathcal{J}$ in Eq.~(\ref{eq:J}) involves two sets of variables: the consensus representations $U$ and the adaptive view weights $\{\lambda^{(v)}\}_{v=1}^{V}$. Directly solving for both simultaneously is difficult. Therefore, we employ an alternating optimization strategy, where we iteratively update one set of variables while keeping the other fixed, until convergence.

\subsubsection{\textbf{Update $U$ (fixing $\{\lambda^{(v)}\}_{v=1}^{V}$)}} 
When the view weights $\{\lambda^{(v)}\}$ are fixed, the objective function with respect to $U$ becomes
\begin{equation}
    \mathcal{J}_{U} = \sum_{v=1}^{V}\lambda^{(v)}\mathcal{L}_{C}^{(v)}(U) + \alpha\mathcal{L}_{Q}(U) + \beta\mathcal{L}_{BB}(U).
\end{equation}
The consensus representations $U$ are updated using a gradient-based optimizer (\textit{e.g.}, Adam~\cite{kinga2015method}). The gradient $\frac{\partial \mathcal{J}_{U}}{\partial U}$ is
\begin{equation}
    \frac{\partial \mathcal{L}_{U}}{\partial U} = \sum_{v=1}^{V}\lambda^{(v)}\frac{\partial \mathcal{L}_{C}^{(v)}(U)}{\partial U} + \alpha\frac{\partial \mathcal{L}_{Q}(U)}{\partial U} + \beta\frac{\partial \mathcal{L}_{BB}(U)}{\partial U}.
\label{eq:totalloss}
\end{equation}

The gradients for quantization loss and bit balance loss are straightforward. Specifically, adapting Eq.~(\ref{eq:Lq}) to be an average value, $\mathcal{L}_{Q}(U)$ becomes
\begin{equation}
    \mathcal{L}_{Q}(U) = \frac{1}{NK} \sum_{i=1}^{N}\sum_{k=1}^{K}(U_{ik}-sign(U_{ik}))^{2}.
\end{equation}
Its derivative is calculated as
\begin{equation}
     \frac{\partial \mathcal{L}_{Q}(U)}{\partial U_{ik}} = \frac{2}{NK} (U_{ik} - \text{sign}(U_{ik})), 
\end{equation}
where $\text{sign}(U_{ik})$ is treated as a constant during differentiation.
Then, adapting Eq.~(\ref{eq:Lbb}) to be an average, $\mathcal{L}_{BB}(U)$ becomes
\begin{equation}
    \mathcal{L}_{BB}(U) = \frac{1}{K}\sum_{k=1}^{K}\left(\frac{1}{N}\sum_{i=1}^{N}U_{ik}\right)^{2}.
\end{equation}
Its derivative is calculated as
\begin{equation}
    \frac{\partial \mathcal{L}_{BB}(U)}{\partial U_{ik}} = \frac{2}{NK} \left( \frac{1}{N} \sum_{i=1}^{N} U_{ik} \right).
\end{equation}

The gradient of the contrastive multi-view graph loss $\frac{\partial \mathcal{L}_{C}^{(v)}(U)}{\partial U}$ with respect to a specific node embedding $U_k$ (the $k$-th row of $U$) is derived from Eq.~(\ref{eq:Lc}). Considering only the terms where $U_k$ is the anchor node (\textit{i.e.}, $i=k$), the derivative is 
\begin{equation}
    \frac{\partial \mathcal{L}_{C}^{(v)}(U)}{\partial U_k} = \frac{\partial}{\partial U_k} \left( -\sum_{j\in\mathcal{N}_{k}^{v}}\log\frac{\exp(sim(U_{k},U_{j})/\tau)}{\sum_{p\ne k}^{N}\exp(sim(U_{k},U_{p})/\tau)} \right).
\label{eq:gLc}
\end{equation}

Let $s_{xy} = sim(U_x, U_y)$ denote the similarity measurement between $U_x$ and $U_y$, the gradient becomes
\begin{equation}
    \frac{\partial \mathcal{L}_{C}^{(v)}(U)}{\partial U_k} = \frac{1}{\tau} \sum_{j \in \mathcal{N}_k^v} \left( \left( \sum_{p \neq k} P_{kp} \frac{\partial s_{kp}}{\partial U_k} \right) - \frac{\partial s_{kj}}{\partial U_k} \right),
\end{equation}
where $P_{kp} = \frac{\exp(s_{kp}/\tau)}{\sum_{q \neq k} \exp(s_{kq}/\tau)}$. The term $\frac{\partial s_{xy}}{\partial U_x}$ depends on the chosen similarity function $sim(\cdot, \cdot)$.
\begin{itemize}
    \item If $sim(U_x, U_y) = U_x^T U_y$ (dot product): $\frac{\partial s_{xy}}{\partial U_x} = U_y$.
    \item If $sim(U_x, U_y) = \frac{U_x^T U_y}{\|U_x\|_2 \|U_y\|_2}$ (cosine similarity):\\
    Let $\hat{U}_x = U_x / \|U_x\|_2$, we have $\frac{\partial s_{xy}}{\partial U_x} = \frac{1}{\|U_x\|_2} (\hat{U}_y - s_{xy} \hat{U}_x)$.
\end{itemize}
These gradients are then used by the optimizer to update $U$.

\subsubsection{\textbf{Update $\{\lambda^{(v)}\}_{v=1}^{V}$ (fixing $U$)}}
When $U$ is fixed, the objective function with respect to $\{\lambda^{(v)}\}$ becomes
\begin{equation}
    \mathcal{J}_{\lambda} = \sum_{v=1}^{V}\lambda^{(v)}\mathcal{L}_{C}^{(v)}(U) + \eta\sum_{v=1}^{V}(\lambda^{(v)})^{\gamma}.
\end{equation}
By setting the derivation of $\mathcal{J}_{\lambda}$ w.r.t. $\lambda^{v}$ to zero, we get
\begin{equation}
    (\lambda^{(v)})^{\gamma-1} = -\frac{\mathcal{L}_{C}^{(v)}(U)}{\eta \gamma}.
\end{equation}
Then, $\lambda^{v}$ is calculated as
\begin{equation}
    \lambda^{(v)} = \left( -\frac{\mathcal{L}_{C}^{(v)}(U)}{\eta \gamma} \right)^{\frac{1}{\gamma-1}}.
\label{eq:lamda}
\end{equation}

This closed-form solution allows for an efficient update of the view weights. The alternating optimization of $U$ and $\{\lambda^{(v)}\}$ continues until convergence. The complete procedures of CMGHash are summarized in Algorithm~\ref{alg}.

\begin{algorithm}[t]
\caption{Contrastive Multi-view Graph Hashing}
\label{alg:cmghash}
\begin{algorithmic}[1]
\Require Multi-view graph data $\mathcal{G}=\{\mathcal{V}, \mathcal{X}, \mathcal{E}\}$, graph filtering parameters $\{m,s\}$, hash code length $K$, and learning parameters $\{k,\tau,\gamma,\eta,\alpha, \beta,\}$.
\Ensure Binary hash codes $\mathbf{B} \in \{-1, 1\}^{N \times K}$. 
\State $\lambda^{(v)} = 1$;
\State Generate smoothed features $\{S^{(v)}\}$ using Eq.~(\ref{eq:smooth});
\State Construct kNN graph from $S^{(v)}$ to get neighbor sets $\mathbb{N}_i^v$;
\State Initialize parameters of the embedding network;
\While{not convergence}
    \State Compute $\mathcal{L}_{C}^{(v)}(U)$ for each view $v$ using Eq.~(\ref{eq:Lc});
    \State Compute $\mathcal{L}_{Q}(U)$ using Eq.~(\ref{eq:Lq});
    \State Compute $\mathcal{L}_{BB}(U)$ using Eq.~(\ref{eq:Lbb});
    \State Update $U$ in Eq.~(\ref{eq:totalloss}) via Adam;
    \For{each view $v$}
        \State Update $\lambda^{(v)}$ in Eq.~\ref{eq:lamda};
    \EndFor
\EndWhile
\State $\mathbf{B} = \text{sign}(U)$;
\State \Return Binary embeddings $\mathbf{B}$.
\end{algorithmic}
\label{alg}
\end{algorithm}

\section{Experiment}

\begin{table*}[t] 
  \centering
  \caption{The statistical information of multi-view graph datasets used.} 
  \label{tab:dataset_stats_grid} 
  \begin{tabular}{c|c|c|c|c} 
    \toprule
    \textbf{Dataset} & \# \textbf{Nodes} & \# \textbf{Features} & \# \textbf{Graph} and (\# \textbf{Edges}) & \# \textbf{Classes} \\
    \midrule
    ACM & 3,025 & (1,830) & Co-Subject (29,281), Co-Author (2,210,761) & 3 \\
    DBLP & 4,057 & (334) & Co-Author (11,113), Co-Conference (5,000,495), Co-Term (6,776,335) & 4 \\ 
    IMDB & 4,780 & (1,232) & Co-Actor (98,010), Co-Director (21,018) & 3 \\
    Amazon photos & 7,487 & (745),(7,487) & Co-Purchase (119,043) & 8 \\
    Amazon computers & 13,381 & (767),(13,381) & Co-Purchase (245,778) & 10 \\
    \bottomrule
  \end{tabular}
\label{tab:datasets}
\end{table*}

\begin{table*}[t]
  \centering
  \caption{Performance about mAP@all score with respect to different number of bits ($K=$8, 16, and 32) on benchmark datasets. The best and the second-best results are in bold and \underline{underline}, respectively. ``N/A'' indicates that the method ran out of memory.}
  \resizebox{\linewidth}{!}{
  \begin{tabular}{l|ccc|ccc|ccc|ccc|ccc}
    \toprule
    \multicolumn{1}{c|}{\multirow{2}{*}{Method}} & \multicolumn{3}{c|}{ACM} & \multicolumn{3}{c|}{DBLP} & \multicolumn{3}{c|}{IMDB} & \multicolumn{3}{c|}{Amazon photos} & \multicolumn{3}{c}{Amazon computers}  \\
    \cmidrule(lr){2-16}
    & 8bits & 16bits & 32bits & 8bits & 16bits & 32bits & 8bits & 16bits & 32bits & 8bits & 16bits & 32bits & 8bits & 16bits & 32bits \\ \midrule
    GCNH & 0.387 & 0.447 & 0.465 & 0.410 & 0.469 & 0.474 & 0.276 & 0.304 & 0.388 & 0.213 & 0.266 & 0.269 & 0.175 & 0.192 & 0.200\\
    HashGNN & \underline{0.477} & 0.529 & \underline{0.571} & 0.442 & 0.497 & 0.520 & 0.297 & 0.356 & 0.399 & 0.297 & \underline{0.315} & 0.329 & \underline{0.288} & 0.276 & 0.301 \\ \midrule
    JMVFG &  0.419&  0.510&  0.523&  0.334&  0.344&  0.351&  0.392&  0.393& 0.394&  0.169&  0.188&  0.209&  0.201&  0.213& 0.218\\ 
    LHGN &  0.354&  \underline{0.546}&  N/A&  \underline{0.590}&  0.567&  N/A&  0.418&  0.421&  N/A&  0.210&  N/A&  N/A&  N/A&  N/A &N/A\\ \midrule
    MvDH & 0.212 & 0.275 & 0.293 & 0.201 & 0.234 & 0.229 & 0.264 & 0.307 & 0.333 & 0.095 & 0.087 & 0.086 & 0.107 & 0.092 & 0.092 \\
    D-MVE-Hash & 0.395 & 0.436 & 0.451 & 0.414 & 0.432 & 0.501 & \underline{0.462} & \underline{0.517} & \underline{0.509} & 0.266 & 0.294 & 0.297 & N/A & N/A & N/A\\ 
    UKMFS & 0.427 & 0.511  & 0.534 & 0.514 & \underline{0.588} & \underline{0.629} & 0.425 & 0.479 & 0.472 & \underline{0.337} & 0.312 & \underline{0.340} & 0.281 & \underline{0.304} & \underline{0.311} \\ \midrule
    CMGHash & \textbf{0.793} & \textbf{0.832}  & \textbf{0.844}  & \textbf{0.678} & \textbf{0.703} & \textbf{0.731} & \textbf{0.498} & \textbf{0.533}  & \textbf{0.521} & \textbf{0.412} & \textbf{0.460}  & \textbf{0.471} & \textbf{0.358} & \textbf{0.396}  & \textbf{0.421} \\ 
    \bottomrule
  \end{tabular}}
\label{tab:main}
\end{table*}

\subsection{Experiment Setup}

\subsubsection{Benchmarks}
We evaluate the proposed CMGHash on five multi-view graph datasets, \textbf{ACM}, \textbf{DBLP}, \textbf{IMDB}~\cite{fan2020one2multi}, \textbf{Amazon photos} and \textbf{Amazon computers}~\cite{shchur2018pitfalls,pan2021multi}.
The ACM dataset is a paper network where node attributes are bag-of-words representations of paper keywords, and its two graphs are constructed from "Co-Author" and "Co-Subject" relationships. The DBLP dataset is an author network, also with node attributes as bag-of-words of author keywords, featuring three graphs derived from "Co-Author", "Co-Conference", and "Co-Term" relationships. The IMDB dataset is a movie network where node attributes are bag-of-words representations for each movie, and graphs are constructed based on co-actor and co-director relationships. Finally, the Amazon photos and Amazon computers datasets are segments of the Amazon co-purchase network; in these, nodes represent goods, features are bag-of-words from product reviews, and edges signify that goods were purchased together. The statistical information for these five datasets is shown in Table~\ref{tab:datasets}.

\subsubsection{Baselines} We compare the proposed CMGHash against three categories of existing methods, as there are currently no direct competitors for multi-view graph hashing. These categories include:
\begin{enumerate}
    \item \textbf{Single-view graph hashing methods.} These methods are designed for hashing data characterized by a single graph structure and a single set of node attributes. The selected baselines include: \textbf{GCNH}~\cite{zhou2018graph}, which uses graph convolutional network to yield similarity-preserving binary embeddings, and \textbf{HashGNN}~\cite{tan2020learning}, which use a GNN encoder for node representations and a hash layer to encode these representations into hash codes. For these two baselines, we report their results on the best-performing single graph structure and single set of node attributes within the dataset.
    \item \textbf{Multi-view graph learning methods.} This category encompasses techniques that learn continuous representations from multi-view graph data, aiming to fuse information from multiple attribute sets or varied graph relations. Since these methods output real-valued embeddings rather than binary codes, we apply a subsequent binarization step to obtain hash codes. The selected baselines include: \textbf{JMVFG}~\cite{fang2023joint}, which jointly performs multi-view feature selection and graph learning by formulating feature selection with an orthogonal decomposition into view-specific bases and bridges this with multi-view graph learning, \textbf{LHGN}~\cite{LHGN}, a supervised learning method, which combines neighborhood and view constraints to learn consensus representations.

    \item \textbf{Multi-view hashing methods.} These methods are designed to fuse information from multiple data sources into compact binary codes but typically assume that each view only consists of vector-based attribute features rather than graph structures. To apply these to our multi-view graph data, we treat each graph view as an attribute view. The selected baselines include: \textbf{MvDH}~\cite{shen2018multiview}, which performs matrix factorization to generate hash codes as latent representations shared by multiple views, \textbf{D-MVE-Hash}~\cite{yan2020deep}, which uses a view stability evaluation method to explore relationships among views and various data fusion methods in Hamming space, and \textbf{UKMFS}~\cite{hu2025unsupervised}, which aims to identify robust consistent graph representation across views and leverages binary hashing codes to guide feature selection.
\end{enumerate}

These selected methods are representative of their respective categories and also include the latest state-of-the-art approaches, ensuring an effective evaluation for the performance of CMGHash.

\subsubsection{Evaluation metrics} We adopt the mean Average Precision (mAP@all), the most commonly used metric in assessing hashing-based representation methods, for evaluation~\cite{wang2017survey,luo2023survey}. For the mAP calculation, the ground truth is determined by the class labels provided with each dataset; two nodes are considered true neighbors if they belong to the same class.

\begin{table*}[t]
  \centering
  \caption{Ablation study in terms of mAP@all score on benchmark datasets. The best and the second-best results are in bold and \underline{underline}, respectively.}
  \resizebox{\linewidth}{!}{
  \begin{tabular}{l|ccc|ccc|ccc|ccc|ccc}
    \toprule
    \multicolumn{1}{c|}{\multirow{2}{*}{Method}} & \multicolumn{3}{c|}{ACM} & \multicolumn{3}{c|}{DBLP} & \multicolumn{3}{c|}{IMDB} & \multicolumn{3}{c|}{Amazon photos} & \multicolumn{3}{c}{Amazon computers}  \\
    \cmidrule(lr){2-16}
    & 8bits & 16bits & 32bits & 8bits & 16bits & 32bits & 8bits & 16bits & 32bits & 8bits & 16bits & 32bits & 8bits & 16bits & 32bits \\ \midrule
    CMGHash-f & 0.713 & 0.742 & 0.760 & 0.630 & 0.656 & 0.667 & 0.441 & 0.474 & 0.503 & 0.366 & 0.428 & 0.439  & 0.311 & 0.363 & 0.382 \\
    CMGHash-q & 0.749 & 0.782 & 0.798 & 0.642 & \underline{0.691} & \underline{0.718} & 0.487 & \underline{0.524} & \textbf{0.525} & 0.407 & 0.450 & 0.453 & 0.337 & 0.380 & 0.407 \\
    CMGHash-b & \underline{0.776} & \underline{0.805} & \underline{0.829} & \underline{0.653} & 0.688 & 0.702 & \underline{0.492} & 0.511 & 0.509 & \textbf{0.417} & \underline{0.452} & \underline{0.462} & \underline{0.357} & \underline{0.388}  & \underline{0.413} \\ 
    CMGHash & \textbf{0.793} & \textbf{0.832}  & \textbf{0.844}  & \textbf{0.678} & \textbf{0.703} & \textbf{0.731} & \textbf{0.498} & \textbf{0.533}  & \underline{0.521} & \underline{0.412} & \textbf{0.460}  & \textbf{0.471} & \textbf{0.358} & \textbf{0.396}  & \textbf{0.421} \\ 
    \bottomrule
  \end{tabular}}
\label{tab:as}
\end{table*}

\begin{figure*}[h!] 
    \centering 
    \begin{subfigure}[b]{0.22\textwidth} 
        \centering
        \includegraphics[width=\textwidth]{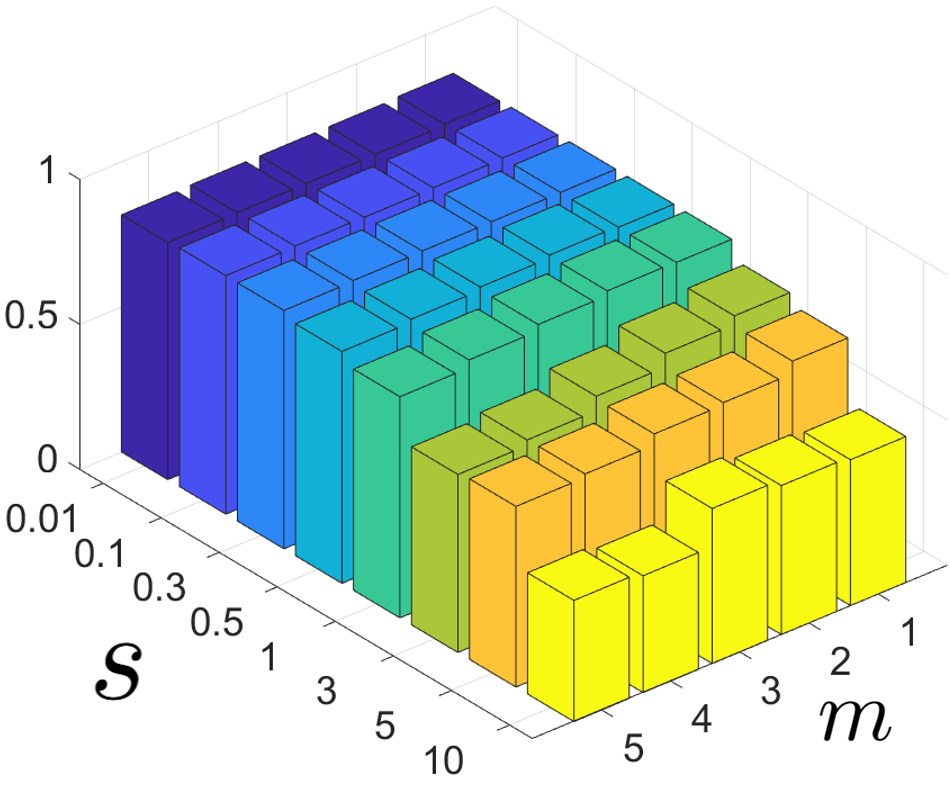} 
        \caption{$s$ and $m$.}
        \label{fig:subfig_a}
    \end{subfigure}
    \hfill 
    \begin{subfigure}[b]{0.22\textwidth}
        \centering
        \includegraphics[width=\textwidth]{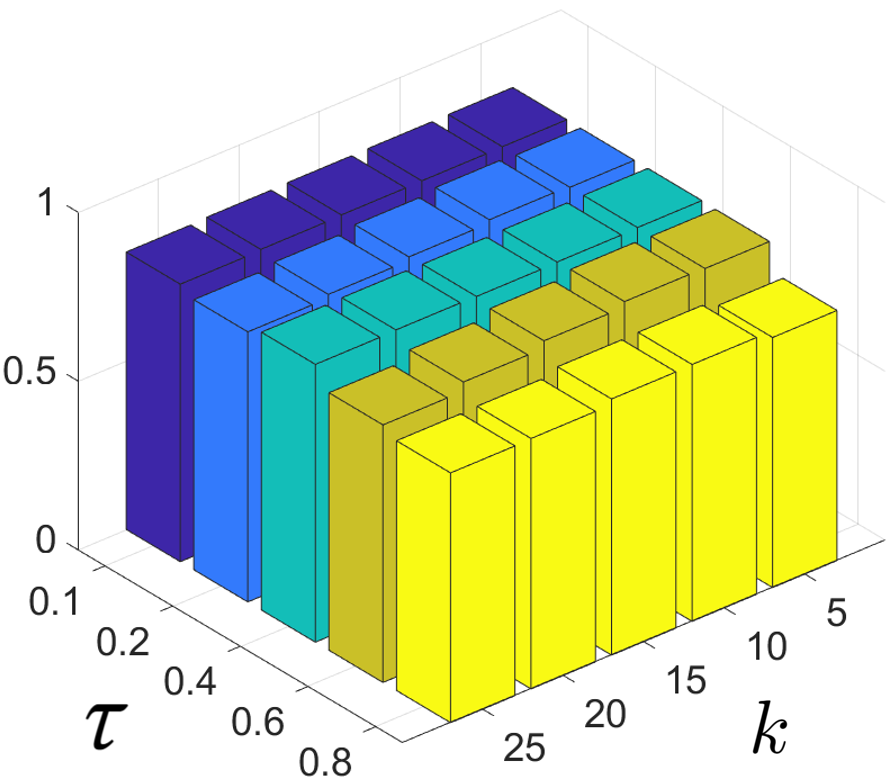} 
        \caption{$\tau$ and $k$.}
        \label{fig:subfig_b}
    \end{subfigure}
    \hfill 
    \begin{subfigure}[b]{0.22\textwidth}
        \centering
        \includegraphics[width=\textwidth]{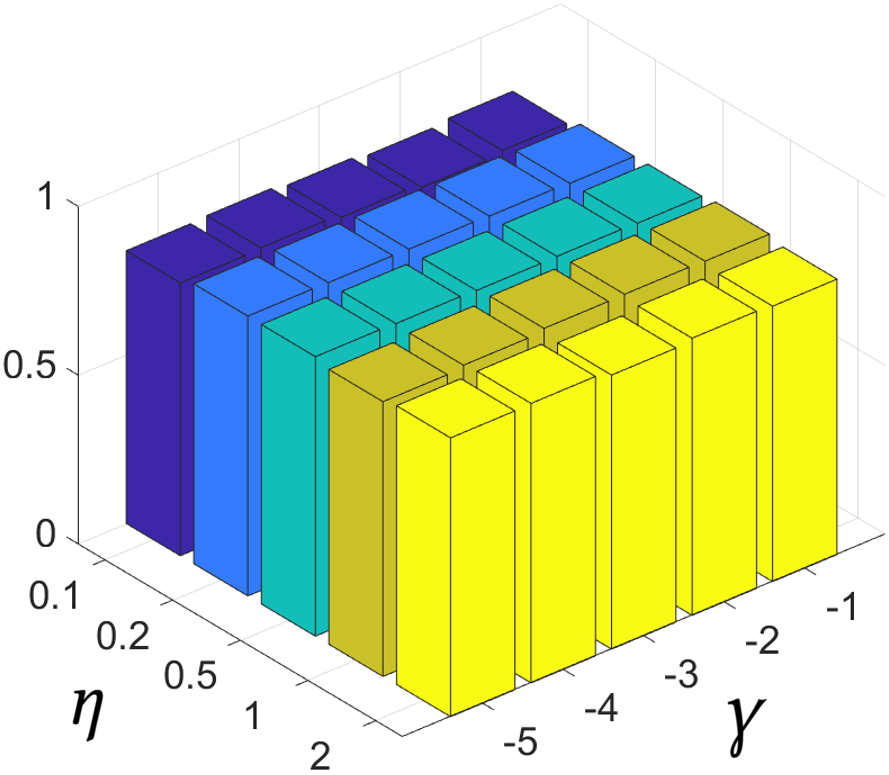} 
        \caption{$\eta$ and $\gamma$.}
        \label{fig:subfig_c}
    \end{subfigure}
    \hfill 
    \begin{subfigure}[b]{0.22\textwidth}
        \centering
        \includegraphics[width=\textwidth]{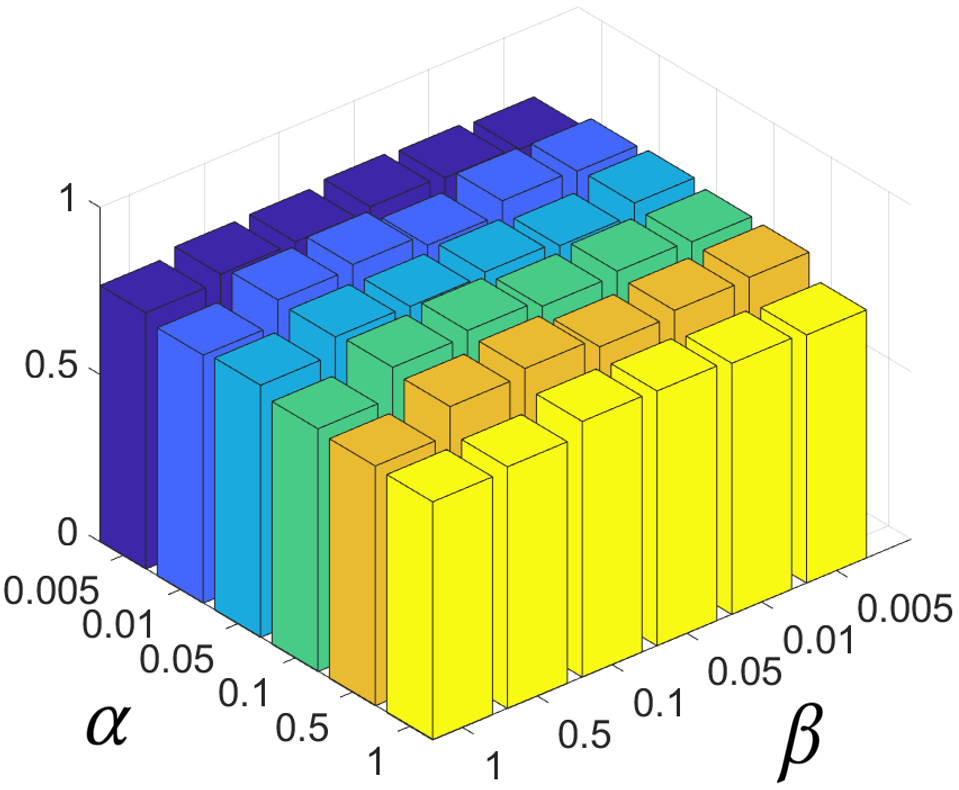} 
        \caption{$\alpha$ and $\beta$.}
        \label{fig:subfig_d}
    \end{subfigure}
    \hfill
    \caption{The parameter sensitivity with mAP@all score about four groups of parameters on the ACM dataset (16 bits).}
    \Description{This figure presents a parameter sensitivity analysis for the CMGHash method through four subplots, each showing performance on the ACM dataset. The first subplot, for graph filtering parameters $s$ and $m$, indicates that good performance is achieved with relatively small values. The second subplot, for contrastive loss parameters $\tau$ and $k$, shows that performance is generally insensitive to $k$ but has an optimal range for $\tau$. The third subplot demonstrates that performance is not sensitive to the weight regularization parameters $\eta$ and $\gamma$. The fourth subplot, for binarization loss weights $\alpha$ and $\beta$, illustrates that achieving optimal results requires careful tuning to balance the objectives.}
    \label{fig:main_figure}
\end{figure*}

\subsubsection{Settings} For all the benchmark methods, we carefully tuned the parameters according to the suggestions in the papers. Some methods are originally designed for image input, including GCNH, MvDH, and D-MVE-Hash, which first extract multi-view features to be used as input. We adapt these methods by directly feeding the original multi-view graph data instead of this feature extraction process. For the proposed CMGHash, we empirically set
$m=2,s=0.5,k=10,\tau=0.2,\gamma=-4$ and $\eta = 1$, since we find that they have little influence to the result. The parameters $\alpha$ and $\beta$ are searched in $\{0.005,0.01,0.05,0.1,0.5,1\}$. All experiments are conducted on the same Linux machine featuring an AMD 64-core CPU with each core running at 2.70GHz and an NVIDIA RTX A6000 GPU with 48GB of RAM.

\subsection{Results}
\label{sec:results}

Table~\ref{tab:main} presents the mAP@all scores of the proposed CMGHash method and the compared methods on five benchmark datasets. CMGHash exhibits significant performance superiority over the competitors on all the datasets. This shows that the proposed CMGHash effectively integrates both the nodes attributes information and the relational information from multiple views. Existing methods of different types achieve suboptimal performance due to certain design limitations. Single-view graph hashing methods, including GCNH and HashGNN, can only integrate attributes and graph structure from a single view, failing to fully utilize the information from multi-view graph data. Multi-view graph learning methods, including JMVFG  and LHGN, do not explicitly impose binarization constraints on the learned representations; consequently, their representations fail to retain sufficient discriminative power when subjected to simple binarization. Multi-view hashing methods, \textit{i.e.} MvDH, D-MVE-Hash and UKMFS, as they are primarily designed for multiple attribute-based views, are not well-suited for learning effective representations from relational graph data, which also leads to suboptimal results. Notably, in handling large-scale graph data, both LHGN and D-MVE-Hash could exhaust available memory, failing to produce results.

\subsection{Ablation Study}

To further validate the contribution of each component in CMGHash, we design three variants. The comparison performance is summarized in Table~\ref{tab:as}. We have the following observations:

\noindent \textbf{CMGHash-f}: To evaluate the graph filtering, we design CMGHash-f, which is the model without graph filtering. A remarkable dropping appears, indicating the necessity of the graph filtering process.

\noindent \textbf{CMGHash-q}: To evaluate the quantization loss $\mathcal{L}_Q$, we design CMGHash-q, which is the model without the quantization loss. A slight drop in CMGHash-q shows the effectiveness of exploring the quantization loss to preserve the information from learned consensus representations.

\noindent \textbf{CMGHash-b}: To evaluate the bit balance loss $\mathcal{L}_{BB}$, we design CMGHash-b, which is the model without the bit balance loss. A slight drop in CMGHash-b shows the effectiveness of enhancing the binary embeddings through constraining the bit balance.

The results collectively demonstrate the integral role of each component in CMGHash. The most significant performance degradation is observed in CMGHash-f, underscoring the critical importance of the graph filtering process for denoising and smoothing the features. Furthermore, the noticeable performance drops in CMGHash-q and CMGHash-b validate the necessity of the quantization and bit balance losses, which effectively guide the model to learn more discriminative and informative binary codes.

\subsection{Parameter Analyses}

To understand the impact of different hyperparameters on CMGHash, we conducted a sensitivity analysis on the ACM dataset, grouping the parameters into four sets as illustrated in Figure~\ref{fig:main_figure}. 

For the graph filtering parameters ($m$ and $s$), which control the graph filtering process, CMGHash achieves good performance with relatively small values for both $m$ and $s$. Therefore, we set the default values to $m=2$ and $s=0.5$ in our experiment. For the contrastive loss parameters ($k$ and $\tau$), the results suggest that the performance is generally not highly sensitive to $k$ across a range of values, while $\tau$ shows an optimal range, with performance peaking around $\tau=0.2$ or $\tau=0.4$. We set the default values to $k=10$ and $\tau=0.2$. For the weight regularization parameters ($\gamma$ and $\eta$), the results suggest that the performance is not sensitive to both $\gamma$ and $\eta$. We set the default values to $\gamma=-4$ and $\eta=1$.
For the binarization loss weights ($\alpha$ and $\beta$), the results show that achieving optimal results requires careful tuning of $\alpha$ and $\beta$ to appropriately balance the objectives. In our experiments, these were searched within the set $\{0.005, 0.01, 0.05, 0.1, 0.5, 1\}$.

For other datasets, the parameter sensitivity trends are consistent with
those observed for the ACM dataset. Due to space limitations, we omit
a detailed presentation for each.

\section{Discussion}

Our work is primarily inspired by the advancements in multi-view graph clustering, particularly the framework of MCGC~\cite{pan2021multi}. This connection stems from our view that semantic hashing and clustering, despite their distinct outputs, are deeply related unsupervised learning tasks built upon a shared foundation of representation learning. Conceptually, semantic hashing aims to map semantically similar instances to nearby binary codes in Hamming space. Clustering, in turn, assigns the same discrete label to similar instances, a process which can be viewed as producing a form of ``integer hash code''. Therefore, both tasks fundamentally rely on learning a robust consensus representation that accurately captures the intrinsic similarity structure of the data.
This connection is particularly evident in computer vision, where both hashing and clustering have independently adopted several shared paradigms. These include deep autoencoders for representation learning~\cite{yang2019deep,yang2018semantic}, and contrastive learning~\cite{caron2020unsupervised,ijcai2021p133}. For example, the hashing method CIBHash~\cite{ijcai2021p133} and the clustering method SwAV~\cite{caron2020unsupervised} both operate on the same principle of learning features that are invariant to data augmentations. The key distinction lies in their final objective: CIBHash tailors this principle to produce discrete binary codes by tackling non-differentiability, while SwAV uses it to generate a continuous feature space suitable for grouping into clusters.

Our work lies in designing a framework that adapts the powerful multi-view graph contrastive learning paradigm specifically for the unique demands of semantic hashing. While the foundational goal of fusing multi-view graph information to learn a consensus representation is shared with clustering, our work introduces critical, task-specific components. This involves not only the $k$NN-based contrastive strategy designed to fuse various structural semantics, but also the incorporation of explicit binarization constraints ($\mathcal{L}_{Q}$ and $\mathcal{L}_{BB}$). These constraints are essential for bridging the gap between the continuous representation space and the discrete Hamming space, a challenge central to hashing but not to clustering. In fact, we argue that the core of a robust framework for learning consistent representations from multi-view graph data relies heavily on how to learn consensus representations rather than being limited by downstream tasks. A paradigm for learning consensus representations is largely universal across tasks such as clustering and semantic hashing.

\section{Conclusion}

In this work, we uncover that existing approaches for mapping multi-view graph data to binary embeddings all suffer from specific limitations. To overcome these, we propose CMGHash, a novel end-to-end framework for multi-view graph hashing. CMGHash first employs graph filtering for smoothed representations. It then learns a unified low-dimensional consensus representation space guided by a $k$NN-based multi-view graph contrastive loss. Moreover, to bridge the gap between continuous representations and binary codes, CMGHash incorporates explicit binarization constraints, facilitating the conversion to binary embeddings with minimal information degradation. Our extensive experiments conducted on five benchmark multi-view graph datasets show that CMGHash significantly outperforms existing approaches. 
As far as we know, CMGHash is the first hashing method specifically designed for multi-view graph data, which effectively integrates node attributes and relational graphs from multiple views, providing new insights for multi-view graph representations. Future research shall focus on leveraging multi-view graph hashing techniques to achieve efficient retrieval and various downstream tasks on large-scale graph data.

\begin{acks}
This work was supported in part by the National Natural Science Foundation of China (Grant No. 92470116).
\end{acks}

\bibliographystyle{ACM-Reference-Format}
\bibliography{bibsample}

\appendix

\end{document}